\documentclass{jjap3}

\title{Multi Graphene Growth on Lead Pencil Drawn Sliver-Halide Print Paper\\
Irradiated by Scanning Femtosecond Laser\\}

\author{Satoru Kaneko$^{1,6}$, Yoshitada Shimizu$^{1}$, Takeshi Rachi$^{1}$, Chihiro Kato$^{1}$, Satomi Tanaka$^{1}$, Yasuhiro Naganuma$^{1}$,Toru Katakura$^{2}$, Kazuo Satoh$^{3}$, Mikio Ushiyama$^{4}$, Seiji Konuma$^{4}$, Yuko Itou$^{4}$, Hirofumi Takikawa$^{5}$,  Goon Tan$^{6}$, Akifumi Matsuda$^{6}$ and Mamoru Yoshimoto$^{6}$}

\inst{
$^{1}$Kanagawa Industrial Technology Research Institute,
Ebina, Kanagawa 243-0435 Japan\\
$^{2}$ RDS Platform, SONY Corp, Atsugi, Kanagawa 243-0014 Japan\\
$^{3}$ Technology Research Institute of Osaka Prefecture, Izumi, Osaka 594-1157 Japan\\
$^{4}$ Kanagawa Academy of Science and Technology, Kawasaki, Kanagawa 213-0012 Japan\\
$^{5}$ Toyohashi University of Technology, Toyohashi, Aichi 441-8580 Japan\\
$^{6}$ Tokyo Institute of Technology, Yokohama, Kanagawa 226-8503 Japan\\
}

\abst{
A variety of paper were drawn by lead pencil with the grade between 4H through 10B. Raman spectroscopy verified both G and D peaks on all the drawing on PC print paper, PC photo paper, kent paper and paper for silver halide print. After irradiation of scanning femtosecond laser, silver halide paper drawn with 10B lead pencil remained surface flatness compared to the other papers. Raman spectroscopy on silver print paper showed a large G peak with less intensity of D peak.  After irradiation of scanning femtosecond laser on silver halide paper drawn by 10B lead pencil, Raman spectroscopy showed large G peak and less intensity of D peak together with 2D peak around 2,700 cm$^{-1}$ corresponding to the existence of multi graphene.}

\begin{document}

\maketitle

\section{Introduction}

\begin{figure}[b]
\center
\includegraphics[angle=0, scale=0.6]{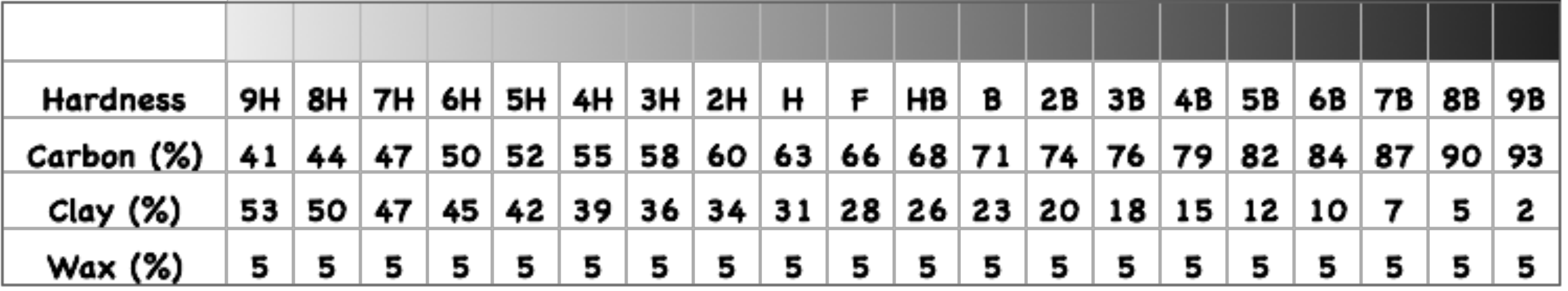}
\caption{\label{fig:Grade_Of_Pencil}(color online) Grade of lead pencil.``H" (for hardness) to``B'' (for blackness), as well as``F'' (Fine), a letter arbitrarily chosen to indicate midway between HB and H.}
\end{figure}

After the discovery of graphene~\cite{novoselov2004,novoselov2005,sato2015} prepared by peeling scotch tape off graphite, fabrication of graphene has been prepared by other methods such as thermal decomposition of SiC~\cite{berger2004,hibino2008,bolen2009}, Chemical Vapor Deposition with catalyst such as Ni, Cu and Fe~\cite{kondo2010,su2011,scott2011,jerng2011,michon2014}. In this study, we propose another method to prepare graphitic carbon by irradiating pencil drawn print paper using femto-second laser.

%
%
Pulsed laser deposition is one of most versatile methods to grow metal~\cite{scheibe1990}, oxides~\cite{kaneko2001,kaneko2004b,kaneko2010a}, multilares~\cite{kaneko2005b,kaneko2006b} including graphene~\cite{cappelli2007,kumar2013} and diamond~\cite{yoshimoto1999}. In these technique, an incident laser beam irradiates a target in a vacuum chamber, and the vaporized materials from the target surface reaches a substrate placed against the target, resulting in growth of film. A femtosecond laser is often used as a laser source~\cite{qian1999,McCann2005}.
We drew a sheet of  printing paper with lead pencil, and irradiated the drawn paper by femtosecond laser
in atmospheric circumstance at room temperature.
Compared to ordinal methods, this simple method using pencil as source material does not required neither high vacuum atmosphere nor high temperature. Vacuum chamber is not required and scanned area is only dependent on movement of X-Y stage (sample holder), resulting in a large advantages of high productivity and large sample size.

The solid core of pencil is made of graphite mixed with clay binder, and is getting``harder" with more clay in the solid core, and soften with more graphite. The ratio of graphite to clay can classify the grade of pencil from ``H'' for hardness to ``B'' for blackness, and another grade ``F'' for fine. A set of pencils ranging from a very hard, light-marking pencil to a very soft, black-marking pencil usually ranges from hardest to softest with different content of carbon, clay and wax~\cite{sousa2001} as shown in Figure~\ref{fig:Grade_Of_Pencil}. The grade of 10B was not available before Mitsubishi Corp. produced the 10B lead pencil for some region of Japan. 

Although paper and pencil are literally primitive materials, an interesting approach was presented~\cite{derman1999,hasan2012,lin2013,kurra2013}. Lin \emph{et. al.} used pencil traces drawn on print papers as strain gauges and as chemiresistors. Graphite particle chains on U-shaped pencil trace vary its resistance with deformation of the paper. As shown in Figure~\ref{fig:Grade_Of_Pencil}, the grade 10B with high graphite content was mainly used for drawing on a sheet of paper in this study. A few kind of papers were also prepared for this study; printing paper for PC printer, photo print paper for PC printer, kent paper and print paper for silver halide prints. The paper for silver halide prints showed lower background fluorescence in Raman spectroscopy, compared to other papers, the grade 10B lead pencil was used for drawing on the backside of paper for sliver halide print (silver print paper) in this study.

\begin{table}[b]
\caption{Irradiation Condition by Femtosecond Laser. The object indicate print paper for PC printer (PC paper), photo print paper for PC printer (PC photo paper), kent paper and paper for silver halide prints (sliver print paper)}
\label{tbl:conditions}
\vspace{10pt}
\begin{center}
\begin{tabular}{cc}  \hline \hline
Object		& PC Paper\\ 
(backside)		& PC Photo Paper\\
			& Kent Paper\\
			& Silver Print Paper\\ \hline
Laser Frequency		& 1 kHz\\ \hline
Wavelength		& 800 nm \\ \hline
Power	&  0 $\sim$ 5 mW\\ \hline
Scanning Speed	& 50 $\sim$ 3,000 $\mu$m/s \\ \hline
Spot Size	&  50 $\mu$m \\ \hline
Pencil	&  9H $\sim$ 10B \\ \hline
\end{tabular}
\end{center}
\end{table}

\section{Experimental}

\begin{figure}
\vspace{10 pt}
\center
\includegraphics[angle=0, scale=0.9]{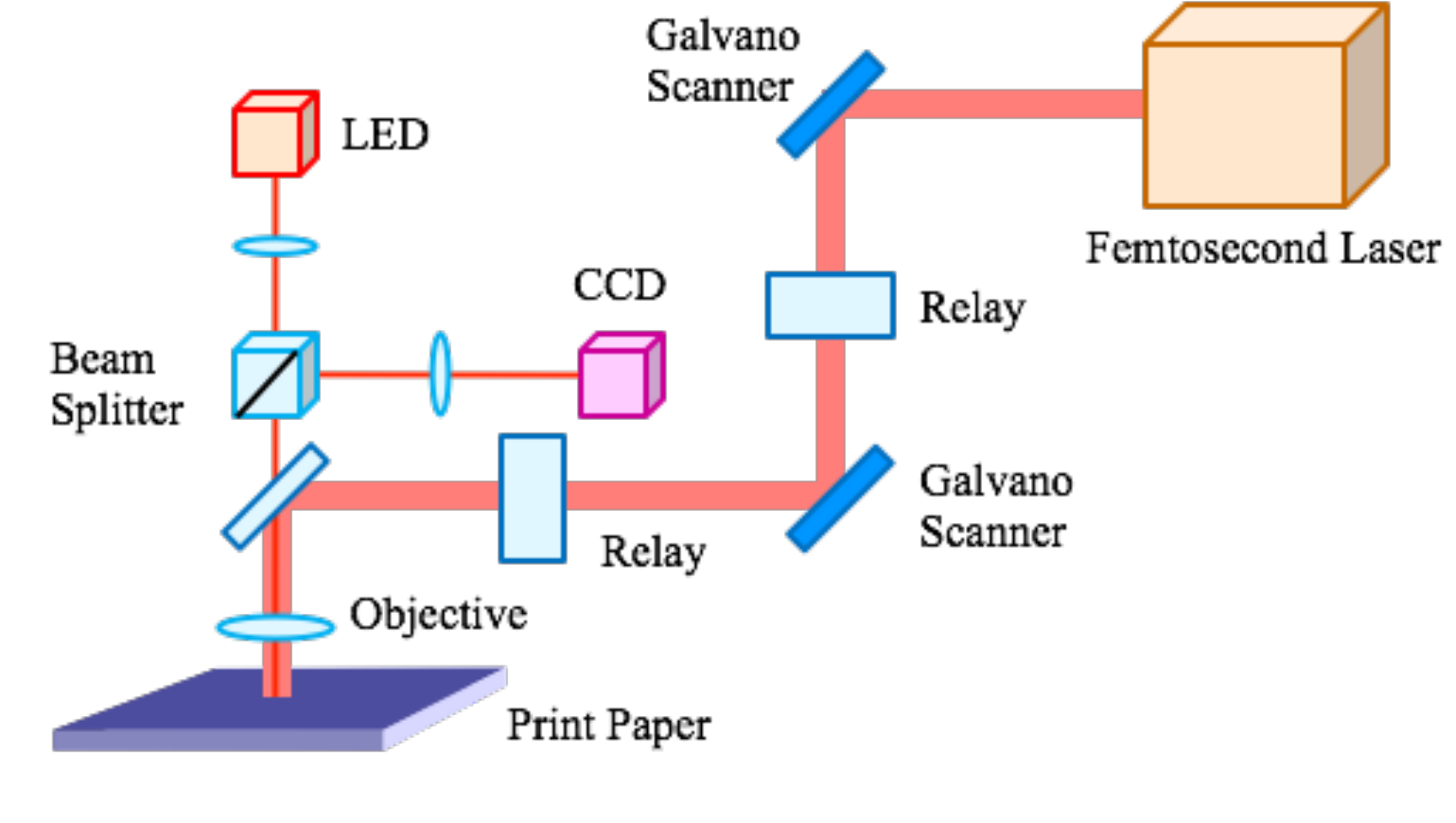}
\caption{\label{fig:Femtosec_Laser}(color online) Schematic of femtosecond laser with wavelength of 800 nm.}
\end{figure}

The papers used in this study were (1) print paper for PC printer (PC paper), (2) photo print paper for PC printer (PC photo paper), (3) kent paper and (4) paper for silver halide prints (sliver print paper). After drawing on paper (roughly size of 10 $\times$ 20 mm) with each grade of lead pencil, Raman spectroscopy with wavelength of 632.8 nm was employed to observe G, D and 2D Raman peaks, corresponding to $\sim$ 1,600, 1,350 and 2,700 cm$^{-1}$, respectively.
Resistance was also measured by multimeter for reference.

A femtosecond laser, (Cyber Laser IFRIT), was employed to irradiate the drawing area with lead pencil on printing paper, as shown in Figure~\ref{fig:Femtosec_Laser}. The max power of laser was 5 mW with wavelength of 800 nm at repetition rate of 1 kHz. The laser spot with diameter of 50 $\mu$m scanned the graphite drawn paper with scan speed between 3,000 to 50 $\mu$m/sec., which resulted in irradiation with laser fluence of between 5 and 0.1 J/cm$^{2}$. A piece of paper fully covered with graphite by pencil drawing was placed on X-Y stage, and the femtosecaond laser was scanned with a line pitch of 25 $\mu$m. The details of irradiation conditions were shown in Table~\ref{tbl:conditions}.

\begin{figure}
\center
\includegraphics[angle=0, scale=0.6]{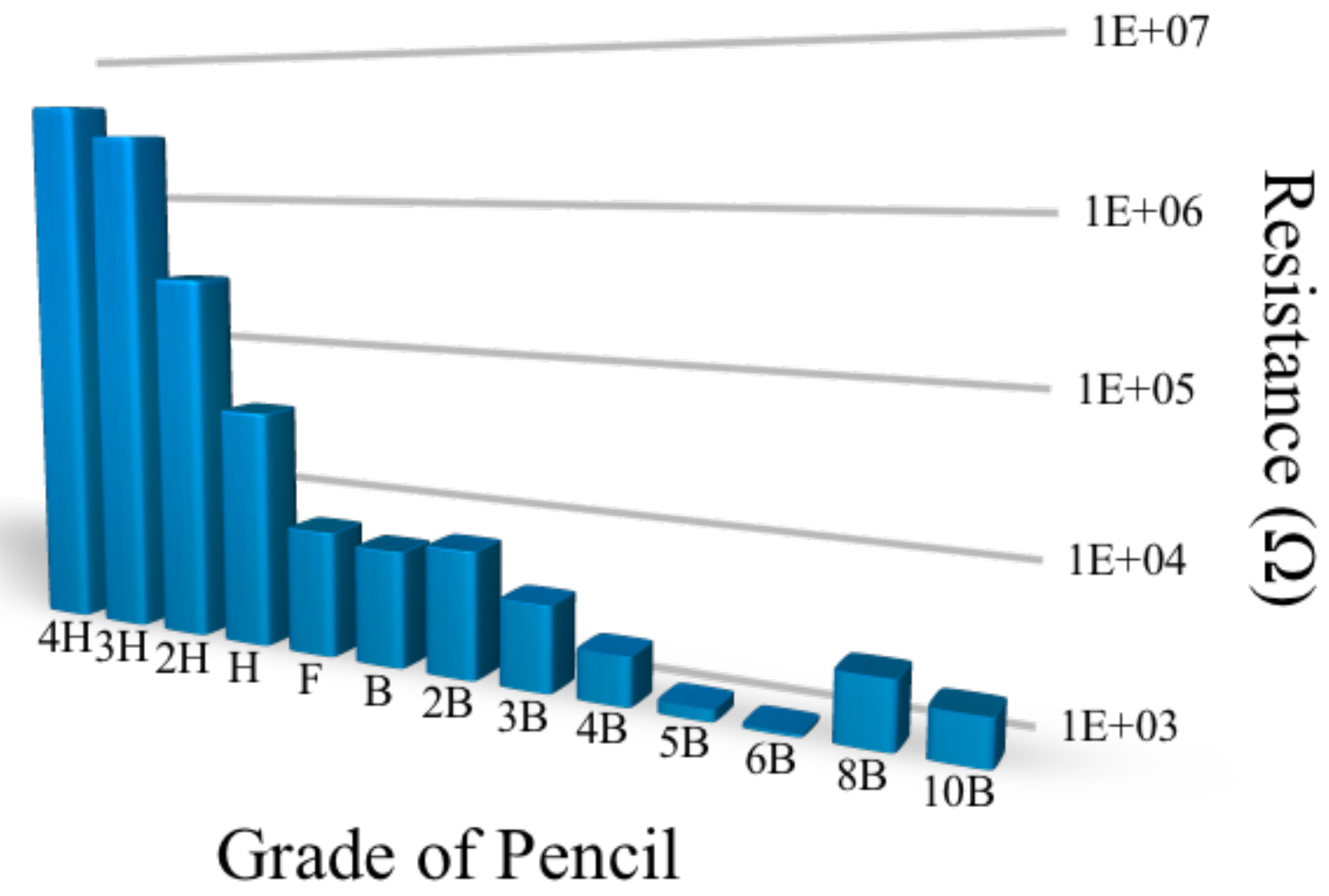}
\caption{\label{fig:Resistance_Of_Pencil}(color online) Irradiated area on (a) kent paper and (b) silver print paper. While finer structure clearly exist on the kent paper, sliver print paper shows the relatively smooth surface.}
\end{figure}

\section{Results \& Discussions}

All the grade of lead pencils were used to draw and filled a area (approximately 1 x 3 mm), and resistance with a length of $\sim$ 20 mm was evaluated by multimeter, as shown in Figure~\ref{fig:Resistance_Of_Pencil}. Multimeter could not measured the resistance of papers drawn by using the grade harder than 5H.  Although the resistances were,  off course,  not accurate and unreliable, they can show tendency depending on the grade of lead pencil. While the grade of 8B and 10B showed little larger resistance than 6B and 4B, those resistance decreased with increase of content of graphite in lead pencil in wide range of the grade between 4H and 6B.
%
%
The resistance of 8B and 10B slightly increased because of ununiformity of carbon thickness by drawing soft pencil. However,
the trend depending on the grade was agreed with previous report~\cite{sousa2001}.

\begin{figure}
\center
\includegraphics[angle=0, scale=0.7]{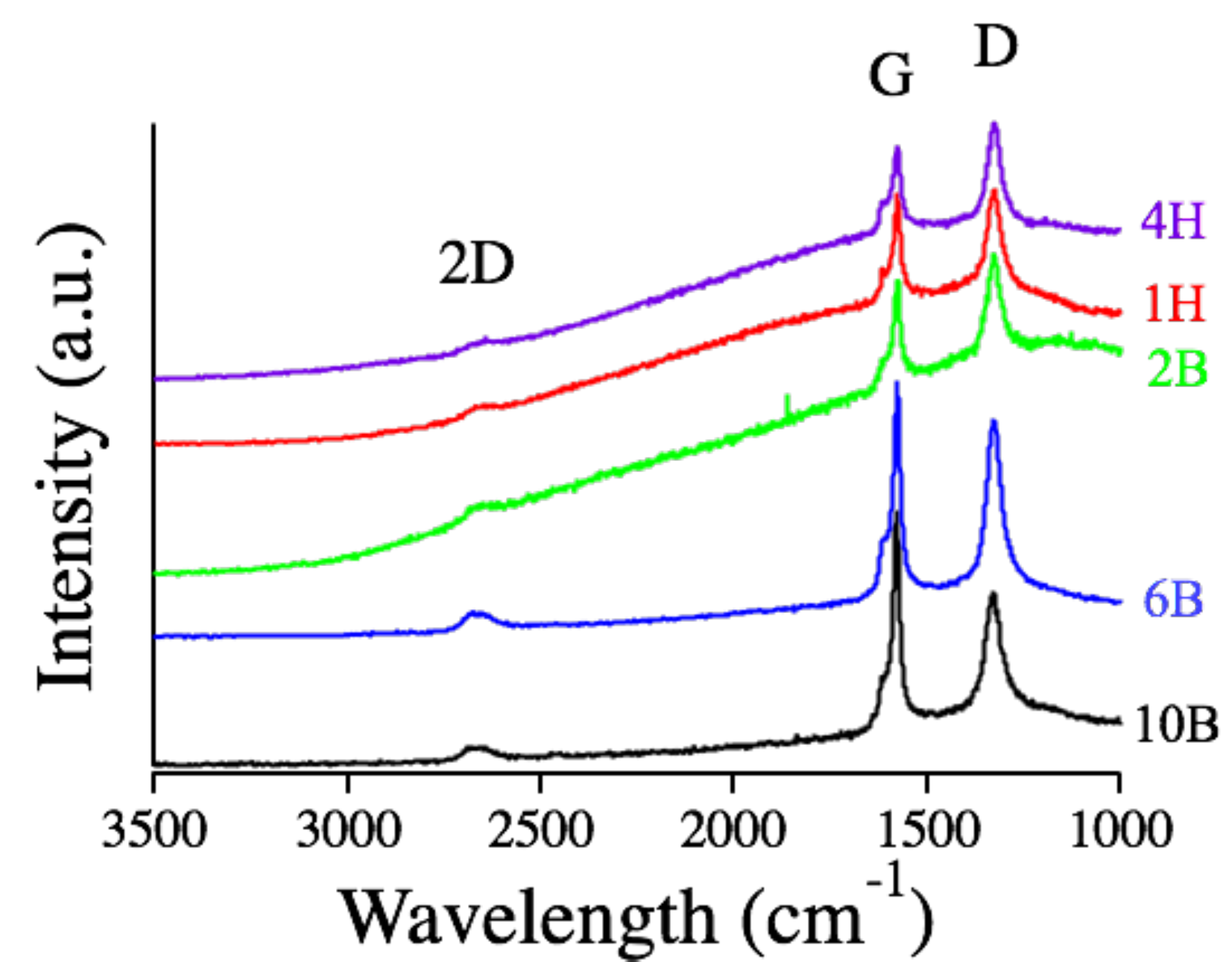}
\caption{\label{fig:Raman_Before_Irradiation}(color online) Raman spectra on sliver print paper drawn by lead pencils with the grade from 4H to 10B.}
\end{figure}

Raman spectra were observed on the drawing area before irradiation of femtosecond laser, as shown in Figure~\ref{fig:Raman_Before_Irradiation}. Although there existed large fluorescence around 1,500 cm$^{-1}$ on harder lead pencil, such as 4H and 1H, all the Raman spectra showed both G and D peaks, corresponding sp$^{2}$ hybridisation (graphitic signature of carbon) and disorder due to defects~\cite{ferrari2006,gupta2006}. It is hard to recognise Raman peak of even D \& G on lead pencil harder than 6H because of thinner drawing and fluorescence of print paper. While the grade between 4H and 2B showed a larger intensity of D peak than G peak with a large background fluorescence, 10B and 6B showed clear Raman peaks with large G peaks more than intensity of D peak. The grade of 10B lead pencil was used hereafter in this study .

It should be noted that the excitation wavelength of 632.8 nm was used with Raman spectroscopy. It is common that 2D peak corresponding to existence of graphene become stronger in Raman spectra with stimulated wavelength of 532 nm. National Institute of Science and Technology (NIST) provides Standard Reference Materials (SRM) to correct the relative intensity of Raman spectra obtained with different excitation wavelength.~\cite{choquette2007} However SRMs 2241 through 2243 provide by NIST can not correct spectra obtained by wavelength of 632.8 nm. The wavelength of 632.8 nm is suitable for graphene on catalyst such as Cu, but it might be unsuitable for one on silver print paper.

\begin{figure}
\center
\includegraphics[angle=0, scale=0.7]{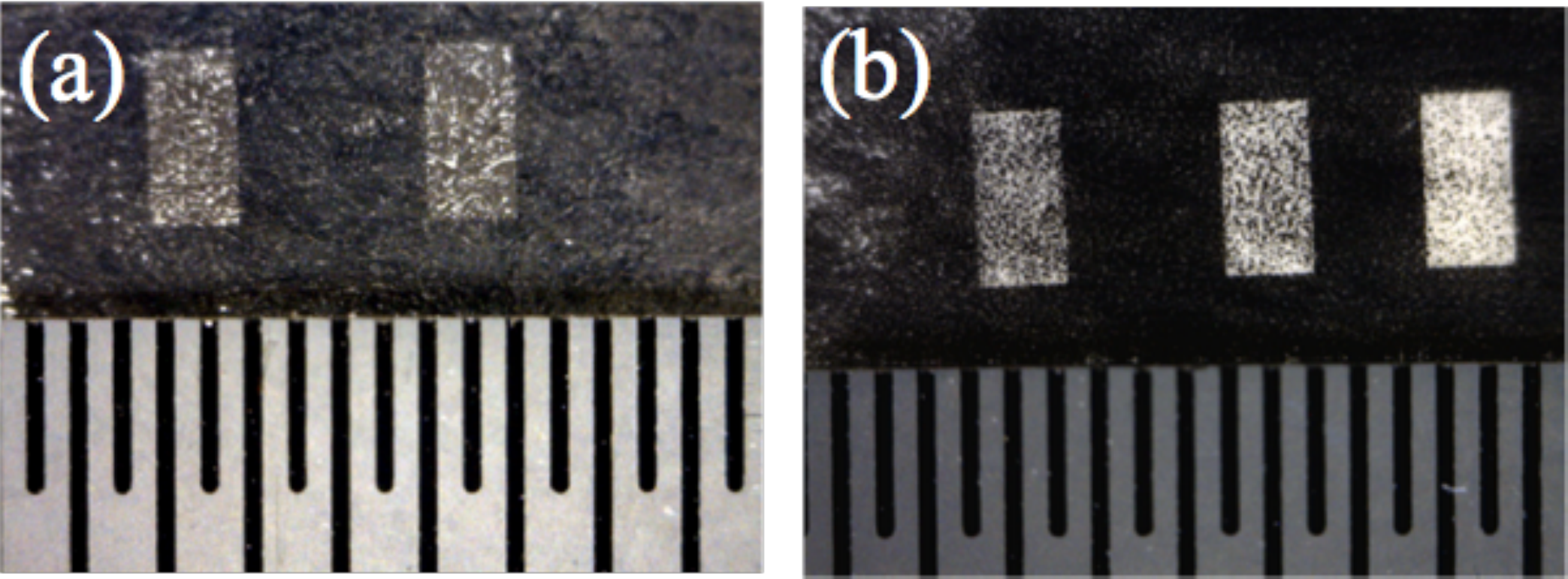}
\caption{\label{fig:Paper_After_Irradiation}(color online) Irradiated area on (a) kent paper and (b) silver print paper. While finer structure clearly exist on the kent paper, sliver print paper shows the relatively smooth surface.}
\end{figure}

\begin{figure}
\center
\includegraphics[angle=0, scale=0.7]{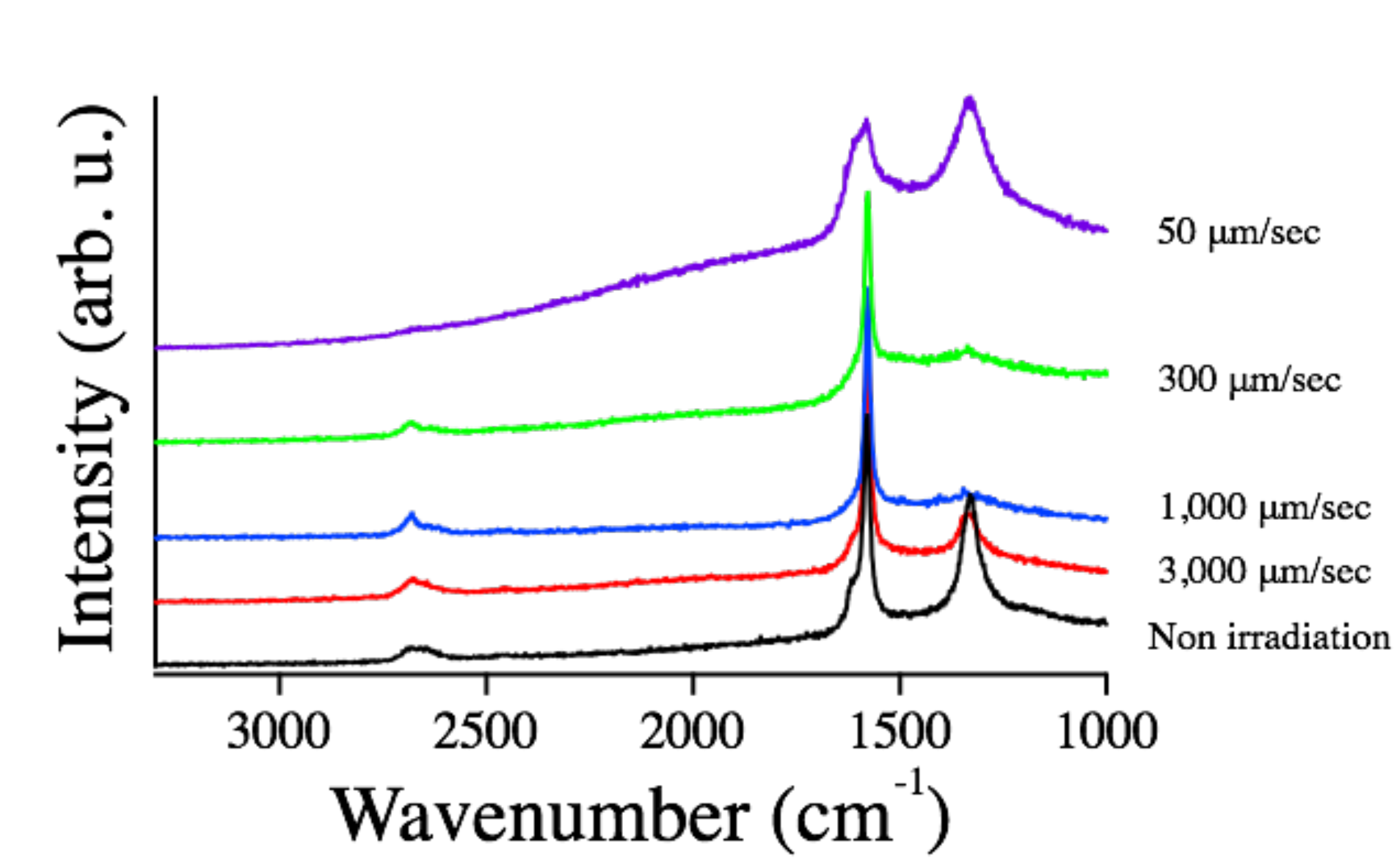}
\caption{\label{fig:Raman_Varied_Speed}(color online) Raman spectra on sliver print paper drawn by 10B lead pencils after irradiation of femtosecond laser with scanning speed up to 3,000 $\mu$m/sec.}
\end{figure}

After filling the area of around 10 $\times$ 20 mm$^{2}$ with 10B pencil, femtosecond laser was employed to irradiate those papers. With high fluency and power, PC paper became scorch and burnt including PC photo paper. A high polymer in PC glossy paper can be easily reacted by laser irradiation. Figure~\ref{fig:Paper_After_Irradiation} shows irradiated area on (a) kent paper and (b) silver print paper. While finer structure clearly exist on the kent paper, sliver print paper shows the relatively smooth surface. Silver print paper was only used hereafter for this study.

Both laser power and scan speed of femtosecond laser were varied, and Raman spectra was examined to optimise the irradiation conditions. Figure~\ref{fig:Raman_Varied_Speed} shows Raman spectra of samples prepared with scan speed in range between 0 to 3,000 $\mu$m/sec. at laser power of 5 mW. At speed of 50 $\mu$m/sec., the excess power removed almost all the graphite drawn on silver print paper, and remained only G and D peaks on Raman spectra.
%
%
With laser power less than 5 mW, the Raman peak D remained at the laser scan speed of 1,000 $\mu$m/sec.
The irradiation with 5 mW at scan speed of 1,000 $\mu$m/sec eliminated D peak and remained G peak together with 2D peak corresponding to growth of graphene.

%
%
Pencil lead consists of graphite, clay and wax, as shown in Table~\ref{fig:Grade_Of_Pencil}. 
We assumed the laser irradiation resulted in  (1) removing impurity other than content of graphite, and (2) promoting growth of graphene. 
%
%
The resistance, for instance, decreased by 10 \% after laser irradiation.
The irradiated area became depressed because the thickness of graphite layer on print paper became thiner after the laser irradiation. Even the resistance decreased by only 10\%, the resistivity can be much improved because of decreasing the graphite thickness. However, the graphite thickness has not been clearly defined on pencil drawing paper. Since the layer thickness is used to evaluate mobility of the graphite layer, it is required to find the way to evaluate thickness of graphite layer drawn on print paper. 


 
In summary, a variety of paper was drawn by lead pencil with the grade between 4H through 10B. Raman spectroscopy verified both G and D peaks, corresponding sp$^{2}$ hybridisation (graphitic signature of carbon) and disorder due to defects. After irradiation of scanning femtosecond laser on silver paper drawn by 10B lead pencil, Raman spectroscopy showed the laser scanning eliminated D peak and remained G and 2D peaks, which indicating the growth of graphene on silver print paper.

\acknowledgement
We would like to thank Yoshitsugu Sato at Kanagawa Industrial Technology Center for technical support. This research was supported in part by Grant- in-Aid for Scientific Research (KAKENHI) A 24246048 and C 26420692.


\begin{thebibliography}{10}
\expandafter\ifx\csname url\endcsname\relax
  \def\url#1{\texttt{#1}}\fi
\expandafter\ifx\csname urlprefix\endcsname\relax\def\urlprefix{URL }\fi

\bibitem{novoselov2004}
K.~S. Novoselov, A.~K. Geim, S.~V. Morozov, D.~Jiang, Y.~Zhang, S.~V. Dubonos,
  I.~V. Grigorieva and A.~A. Firsov: Science  {\bf 306} (2004) 666.

\bibitem{novoselov2005}
K.~S. Novoselov, D.~Jiang, F.~Schedin, T.~J. Booth, V.~V. Khotkevich, S.~V.
  Morozov and A.~K. Geim: PNAS  {\bf 102} (2005) 10451.

\bibitem{sato2015}
S.~Sato: Jpn. J. Appl. Phys.  {\ bf54} (2015) 040102.

\bibitem{berger2004}
C.~Berger, Z.~Song, T.~Li, X.~Li, A.~Y. Ogbazghi, R.~Feng, Z.~Dai, A.~N.
  Marchenkov, E.~H. Conrad, P.~N. First and W.~A. de~Heer: J. Phys. Chem. B  {\bf 108}
  (2004) 19912.

\bibitem{hibino2008}
H.~Hibino, H.~Kageshima, F.~Maeda, M.~Nagase, Y.~Kobayashi and H.~Yamaguchi: Phys.
  Rev. B  {\bf 77} (2008) 075413.

\bibitem{bolen2009}
M.~L. Bolen, S.~E. Harrison, L.~B. Biedermann and M.~A. Capano: Phys. Rev. B  {\bf 80}
  (2009) 115433.

\bibitem{kondo2010}
D.~Kondo, S.~Sato, K.~Yagi, N.~Harada, M.~Sato, M.~Nihei and N.~Yokoyama: Appl.
  Phys. Express  {\bf 3} (2010) 025102.

\bibitem{su2011}
C.-Y. Su, A.-Y. Lu, C.-Y. Wu, Y.-T. Li, K.-K. Liu, W.~Zhang, S.-Y. Lin, Z.-Y.
  Juang, Y.-L. Zhong, F.-R. Chen and L.-J. Li: Nano Lett.  {\bf 11} (2011) 3612.

\bibitem{scott2011}
A.~Scott, A.~Dianat, F.~B{\"o}rrnert, A.~Bachmatiuk, S.~Zhang, J.~H. Warner,
  E.~Borowiak-Palen, M.~Knupfer, B.~Buchner, G.~Cuniberti and M.~H. R{\"u}mmeli:
  Appl. Phys. Lett.  {\bf 98} (2011) 073110.

\bibitem{jerng2011}
S.-K. Jerng, D.~S. Yu, J.~H. Lee, C.~Kim, S.~Yoon and S.-H. Chun: Nano. Res. Lett.
   {\bf 6} (2011) 565.

\bibitem{michon2014}
A.~Michon, A.~Tiberj, S.~Ve􏰀zian, E.~Roudon, D.~Lefebvre, M.~Portail,
  M.~Zielinski, T.~Chassagne, J.~Camassel and Y.~Cordier: Appl. Phys. Lett.  {\bf 104}
  (2014) 071912.

\bibitem{scheibe1990}
H.~J. Scheibe, A.~A. Gorbunov, G.~K. Baranova, N.~V. Klassen, V.~I. Konov and
  M.~P. Kulakov: Thin Solid Films  {\bf 189} (1990) 283.

\bibitem{kaneko2001}
S.~Kaneko, Y.~Shimizu and S.~Ohya: Jpn.\ J.\ Appl.\ Phys.  {\bf 40} (2001) 4870.

\bibitem{kaneko2004b}
S.~Kaneko, Y.~Shimizu, K.~Akiyama, T.~Ito, M.~Mitsuhashi, S.~Ohya, K.~Saito,
  H.~Funakubo and M.~Yoshimoto: Appl.\ Phys.\ Lett.  {\bf 85} (2004) 2301.

\bibitem{kaneko2010a}
S.~Kaneko, T.~Nagano, K.~Akiyama, T.~Ito, M.~Yasui, Y.~Hirabayashi,
  H.~Funakubo and M.~Yoshimoto: J. Appl. Phys.  {\bf 107} (2010) 073523.

\bibitem{kaneko2005b}
S.~Kaneko, K.~Akiyama, Y.~Shimizu, H.~Yuasa, Y.~Hirabayashi, S.~Ohya, K.~Saito,
  H.~Funakubo and M.~Yoshimoto: J. Appl. Phys.  {\bf 97} (2005) 103904.

\bibitem{kaneko2006b}
S.~Kaneko, K.~Akiyama, H.~Funakubo and M.~Yoshimoto: Phys. Rev. B  {\bf 74} (2006)
  054503.

\bibitem{cappelli2007}
E.~Cappelli, S.~Orlando, V.~Morandi, M.~Servidori and C.~Scilletta: J. Phys. Conf.
  Series  {\bf 59} (2007) 616.

\bibitem{kumar2013}
S.~R.~S. Kumar and H.~N. Alshareef: Appl. Phys. Lett.  {\bf 102} (2013) 012110.

\bibitem{yoshimoto1999}
M.~Yoshimoto, K.~Yoshida, H.~Maruta, Y.~Hishitani, H.~Koinuma, S.~Nishio,
  M.~Kakihara and T.~Tachibana: Nature  {\bf 399} (1999) 340.

\bibitem{qian1999}
F.~Qian, V.~Craciun, R.~K. Singh, S.~D. dutta and P.~P. Pronko: J. Appl. Phys.  {\bf 86}
  (1999) 2281.

\bibitem{McCann2005}
R.~McCann, S.~S. Roy, P.~Papakonstantinou, J.~A. McLaughlin and S.~C. Ray: J.
  Appl. Phys.  {\bf 97} (2005) 093522.

\bibitem{sousa2001}
M.~C. Sousa and J.~W. Buchanan: Computer Graphics Forum  {\bf 19} (2001) 27.

\bibitem{derman1999}
S.~Derman and A.~Goykadosh: Phys. Teach.  {\bf 37} (1999) 400.

\bibitem{hasan2012}
K.~ul~Hasan, O.~Nur and M.~Willander: Appl. Phys. Lett.  {\bf 100} (2012) 211104.

\bibitem{lin2013}
C.-W. Lin, Z.~Zhao, J.~Kim and J.~Huang: Scientific Report  {\bf 4} (2013) 3812.

\bibitem{kurra2013}
N.~Kurra, D.~Duttaa and G.~U. Kulkarni: Phys. Chem. Chem. Phys.  {\bf 15} (2013) 8367.

\bibitem{ferrari2006}
A.~C. Ferrari, J.~C. Meyer, V.~Scardaci, C.~Casiraghi, M.~Lazzeri, F.~Mauri,
  S.~Piscanec, D.~Jiang, K.~S. Novoselov, S.~Roth and A.~K. Geim: Phys. Rev. Lett.
  {\bf 97} (2006) 187401.

\bibitem{gupta2006}
A.~Gupta, G.~Chen, P.~Joshi, S.~Tadigadapa and P.~Eklund: Nano Lett.  {\bf 6} (2006) 2667.

\bibitem{choquette2007}
S.~J. Choquette, E.~S. Etz, W.~S. Hurst, D.~H. Blackburn and S.~D. Leigh: Applied
  Spectroscopy  {\bf 61} (2007) 117.

\end{thebibliography}

\end{document}